\documentclass[aps,pre,one,11pt]{revtex4}%
\usepackage{epsfig}
\usepackage{graphicx}
\usepackage{amsmath}
\usepackage{amsfonts}
\usepackage{amssymb}%
\begin{document}
\title{Improved viscosity-concentration equation for emulsions of nearly spherical droplets}
\author{Carlos I. Mendoza$^{1}$ and I. Santamar\'{\i}a-Holek$^{2}$ }
\affiliation{$^{1}$Instituto de Investigaciones en Materiales, Universidad Nacional
Aut\'{o}noma de M\'{e}xico, Apdo. Postal 70-360, 04510 M\'{e}xico, D.F., Mexico}
\affiliation{$^{2}$Facultad de Ciencias, Universidad Nacional Aut\'{o}noma de M\'{e}xico,
Circuito Exterior de Ciudad Universitaria, 04510, M\'{e}xico D. F., Mexico}

\begin{abstract}
We propose an improved viscosity model accounting for experiments of emulsions
of two immiscible liquids at arbitrary volume fractions and low shear rates.
The model is based on a recursive-differential method formulated in terms of
the appropriate scaling variable which emerges from an analysis of excluded
volume effects in the system. This variable, called the effective filling
fraction, incorporates the geometrical information of the system which
determines the maximum packing and reduces to the bare filling fraction for
infinitely diluted emulsions. The agreement of our model for the viscosity
with experiments is remarkable for all the range of volume fractions and
viscosity ratio.

\end{abstract}
\maketitle


\section{Introduction}

Due to the central role that they play in many technological processes, the
rheology of solid-liquid suspensions is a subject for which a large amount of
work has been produced \cite{einstein,krieger,russel,werff,masalova,larson,mendoza}. However, the rheological
properties of emulsions of immiscible liquids have received much less
attention, despite the fact that they are also very important in many
industrial applications \cite{taylor,yaron,choi,phan,pal1,pal2}.

Emulsions present an interesting rheological behavior with characteristics
similar to those of the solid-liquid suspensions. In particular, the relative
viscosity $\eta_{r}\left(  \phi\right)  $ of emulsions of nearly spherical
droplets also diverges at certain critical value of the filling fraction:
$\eta_{r}\left(  \phi\rightarrow\phi_{c}\right)  \rightarrow\infty$. Also
similar is the fact that for very dilute emulsions where the interaction
between neighboring drops is absent, the relative viscosity follows the
Einstein's like relation%
\begin{equation}
\eta_{r}\left(  \phi\right)  =\frac{\eta\left(  \phi\right)  }{\eta_{c}%
}=\left(  1+\frac{1+2.5K}{1+K}\phi\right)  , \label{taylor}%
\end{equation}
where the viscosity ratio $K=\eta_{d}/\eta_{c}$ contains the viscosity
$\eta_{d}$ of the dispersed phase and the viscosity $\eta_{c}$ of the
continuous phase. Equation (\ref{taylor}) has been originally introduced by
Taylor \cite{taylor} and is called Taylor's equation. The difference between
Einstein's and Taylor's expressions lies in the fact that in the later
relation the coefficient multiplying the filling fraction $\phi$ incorporates
information about the nature of the dispersed phase through its viscosity
$\eta_{d}$. However, when the viscosity of the dispersed phase is much larger
than the viscosity of the continuous phase, $K\rightarrow\infty$, one recovers
Einstein's expression $\eta\left(  \phi\right)  =\eta_{c}\left(
1+2.5\phi\right)  $.

Although of fundamental importance, Taylor's equation does not reproduces the
behavior of concentrated emulsions since in this case interactions between
droplets becomes very important. Several efforts have been done in order to
incorporate these interactions \cite{choi,yaron}. Among them we can mention the
viscosity equation developed in Ref. \cite{phan} based on a differential
effective medium procedure%
\begin{equation}
\eta_{r}\left(  \phi\right)  \left(  \frac{2\eta_{r}\left(  \phi\right)
+5K}{2+5K}\right)  ^{3/2}=\left(  1-\phi\right)  ^{-5/2}, \label{phan}%
\end{equation}
that takes into account the effect of the viscosity ratio $K$. A shortcoming
of this expression is that it fails to describe the experimental data
adequately at large concentrations. Pal improved this model by incorporating
"crowding effects" through a critical filling fraction $\phi_{c}$, \cite{pal1}:
\begin{equation}
\eta_{r}\left(  \phi\right)  \left(  \frac{2\eta_{r}\left(  \phi\right)
+5K}{2+5K}\right)  ^{3/2}=\left(  1-\frac{\phi}{\phi_{c}}\right)
^{-(5/2)\phi_{c}}. \label{model2}%
\end{equation}
This model (called model 2) considerable improves the quantification of the
rheological properties of emulsions of two immiscible liquids even at high
filling fractions. However, comparison with experiments shows that it
underestimates the value of the viscosity at intermediate volume fractions
\cite{pal1}. Pal also proposed the following model giving excellent agreement
with experimental data at large volume fractions \cite{pal2}
\begin{equation}
\eta_{r}\left(  \phi\right)  \left(  \frac{2\eta_{r}\left(  \phi\right)
+5K}{2+5K}\right)  ^{3/2}=\left(  1-\frac{\phi}{\phi_{c}}\right)  ^{-5/2},
\label{pal}%
\end{equation}
however, it does not reduce to the correct Taylor's expression at low concentrations.

Having in mind all these considerations, in this article we propose an
improvement to the differential effective medium model that gives the correct
divergence at the critical concentration, produces a better description of
experiments and reduces to the right Taylor's expression at low concentrations.

\section{Improved differential viscosity model: Scaling and excluded volume
effects}

In its original article, Taylor calculated the viscosity of a dilute
suspension of fluid drops under three main assumptions: \emph{i)} The drops
are small enough to maintain a spherical shape due to surface tension,
\emph{ii)} no slipping exists at the interface between the drops and the host
fluid and \emph{iii)} tangential stresses are continuous at the surface of the
drop. Then, using hydrodynamic results for the velocity field out and inside
of the sphere, he obtained the corresponding expressions for the components of
the stress acting across unit area of the spherical surface and imposed on the
tangential components the continuity condition. The normal stress is not
continuous at the surface, fact which is ultimately related to the surface
tension. As a result of these impositions, Taylor obtained the explicit
expressions for the four constants appearing in the formulas giving the
components of the stress tensor. Then, following Einstein's arguments, Taylor
showed that in the case of liquid droplets, the factor $5/2$ that multiplies
the volume fraction $\phi$ in Einstein's relation $\eta_{r}\left(
\phi\right)  =\left[  1+\left(  5/2\right)  \phi\right]  $ must be replaced by
$F=-5/2\left[  \eta_{d}+(2/5)\eta_{c}\right]  /\left[  \eta_{d}+\eta
_{c}\right]  $. This replacement leads to Eq. (\ref{taylor}).

The procedure followed by Taylor is very interesting since it emphasizes two
\emph{different} ingredients entering into the suspended phase correction
$\Delta\eta$ ($=\eta_{r}-1$) in Eq. (\ref{taylor})
\begin{equation}
\Delta\eta\left(  \phi\right)  =\frac{1+2.5K}{1+K}\phi.\label{Deltataylor}%
\end{equation}
The first ingredient is that the boundary conditions at the surface of the
drop determine the stresses applied on it and thus control the contribution
due of a single drop to the total stress of the composite system: particle
plus continuous phase. The second ingredient comes from the consideration of
the contribution of $N$ droplets to the composite system. Following Landau
\cite{landau}, one may argue that the total contribution by the drops
$\overline{\mathbf{\Pi}}_{d}^{V}$ to the total stress tensor comes from an
average over the volume $V$ of the system in the form
\begin{equation}
\overline{\mathbf{\Pi}}_{d}^{V}\simeq\frac{N}{V}\int\mathbf{\Pi}_{d}%
^{(1)}dV,\label{landau22.5}%
\end{equation}
where we represented the single drop contribution to the stress tensor by
$\mathbf{\Pi}_{d}^{(1)}$ which is proportional to $F$ and $a^{3}$ with $a$ the
radius of the drop. The upper $V$ in Eq.(\ref{landau22.5}) stands for the
volume average and the factor $N$ accounts for the contribution of the $N$
independent particles. However, this average is strictly valid only when the
system is made of point particles. This restricts the validity of Eq.
(\ref{landau22.5}) to very diluted systems.

Thus, if one considers that a drop has a volume $V_{d}$ then the average must
be performed over the free volume accessible to the particles, which is
defined by: $V_{free}=V-cNV_{d}$. Here, $V_{d}$ is the volume of one particle
and $c$ is a constant taking into account the fact that the complete free
volume can not be filled with drops. Note that, for different symmetries of
the suspended particles, the value of the constant $c$ will be different. This
is specially important in the case of drops which may change their shape when
subjected to strong shears. Therefore, $c$ contains information about the
maximum packing of drops the system may allocate.

Therefore, if excluded volume effects are taken into account, the suspended
phase contribution to the stress tensor is given by
\begin{equation}
\overline{\mathbf{\Pi}}_{d}^{V_{free}}\simeq\frac{N}{V-cNV_{d}}\int
\mathbf{\Pi}_{d}^{(1)}dV. \label{landau22.5-B}%
\end{equation}
For finite-sized droplets, this relation leads to the result that the Taylor
expression scales with the excluded volume factor ${\phi} /({1-c\phi})$
instead of $\phi$, and thus gives the following expression for the viscosity
of an emulsion
\begin{equation}
\eta_{r}\left(  \phi\right)  =\left[  1+\frac{1+2.5K}{1+K}\left(  \frac{\phi
}{1-c\phi}\right)  \right]  . \label{taylor-excluded}%
\end{equation}
Taylor's formula is then recovered in the limit of very small volume fractions
($\phi\rightarrow0$). Eq. (\ref{taylor-excluded}) explains the experimental
fact that the viscosity diverges at volume fractions $\phi<1$, implying that
this effect is related to the excluded volume effects. Similar arguments used
in the case of solid particles lead to excellent agreement between experiments
and theory, see Ref. \cite{mendoza}.

The central result of the present analysis is Eq. (\ref{taylor-excluded}),
which shows that the correct scaling factor describing the dependence of the
viscosity of an emulsion of finite-sized droplets on the filling fraction is
the \emph{effective} filling fraction
\begin{equation}
\phi_{eff}=\frac{\phi}{1-c\phi}, \label{phieff}%
\end{equation}
where $c$ is a structural or crowding factor that takes into account the
arrangements of the droplets in the emulsion and is given by
\begin{equation}
c=\frac{1-\phi_{c}}{\phi_{c}}, \label{c}%
\end{equation}
where $\phi_{c}$ is the filling fraction at the divergence. The importance of
$\phi_{eff}$ is that it incorporates correlations between particles by taking
into account geometrical information having the characteristic that
$\phi_{eff}\sim\phi$ at low volume fractions and $\phi_{eff}=1$ at maximum
packing. This quantity plays an important role in determining the correct
dependence of the viscosity of the emulsion, as we will show next.

At low volume fractions, the shear viscosity $\eta$ of an emulsion is given by
Eq.(\ref{taylor-excluded}). In order to extend it to larger filling fractions
$\phi$, we will use a differential effective medium approach in which a
concentrated emulsion is obtained from an initial continuous phase by
successively adding infinitesimally small quantities of droplets to the system
until the final volume fraction of the dispersed phase is reached.

The usual implementation of the differential theory to obtain a concentrated
emulsion up to a given stage, consists in treat it as a homogeneous effective
medium of viscosity $\eta\left(  \phi\right)  $ into which we add a quantity
$\Delta\phi$ of new droplets. Then, the new viscosity $\eta\left(  \phi
+\delta\phi\right)  $ is calculated using Taylor's expression to give
\begin{equation}
\eta\left(  \phi+\delta\phi\right)  =\eta\left(  \phi\right)  \left(
1+\frac{\eta\left(  \phi\right)  +2.5\eta_{d}}{\eta\left(  \phi\right)
+\eta_{d}}\Delta\phi\right)  , \label{recursive1}%
\end{equation}
Here, it is important to notice that in order to allocate the new droplets
maintaining constant the volume of the system, one has to remove part of the
effective medium $(1-\Delta\phi)$ in order to allocate the new quantity
($\Delta\phi$) of droplets. Therefore, the new filling fraction is given by
$\phi+\delta\phi=\phi(1-\Delta\phi)+\Delta\phi$, from which one finds
\begin{equation}
\Delta\phi=\frac{\delta\phi}{1-\phi}. \label{deltaphi}%
\end{equation}

A shortcoming of using Eq. (\ref{recursive1}) and the bare filling fraction
$\phi$ is that it does not contains the correlations generated by the excluded
volume effects, that is, it assumes that all the volume of the emulsion before
new droplets are added is available to the new droplets. However, as we have
mentioned previously, this is not possible since the droplets can occupy only
the free volume and therefore scale according to $\phi_{eff}=\phi
(1-c\phi)^{-1}$, Eq. (\ref{phieff}).

These considerations suggest that the differential theory must be performed in
terms of the scaling variable $\phi_{eff}$ and not on the bare filling
fraction itself.

Therefore, making the substitution of $\phi$ by $\phi_{eff}$ in
Eqs.(\ref{recursive1}) and (\ref{deltaphi}) and integrating we obtain
\begin{equation}
\eta_{r}\left(  \phi_{eff}\right)  \left(  \frac{2\eta_{r}\left(  \phi
_{eff}\right)  +5K}{2+5K}\right)  ^{3/2}=\left(  1-\phi_{eff}\right)  ^{-5/2},
\label{viscosity1}%
\end{equation}
that, in terms of the bare filling fraction takes the form
\begin{equation}
\eta_{r}\left(  \phi\right)  \left(  \frac{2\eta_{r}\left(  \phi\right)
+5K}{2+5K}\right)  ^{3/2}=\left(  1-\frac{\phi}{1-c\phi}\right)  ^{-5/2}.
\label{viscosity-bare}%
\end{equation}
Expressions (\ref{viscosity1}) and (\ref{viscosity-bare}) are similar to other
ones based on the differential method, see, for example, Eqs. (\ref{phan}%
)-(\ref{pal}). However, our model crucially makes use of the geometrical
information of the system from the start through $\phi_{eff}$. This makes an
essential difference since, in contrast to the usual differential procedure,
the use of $\phi_{eff}$ as the integration variable incorporates correlations
between droplets of the same recursive stage, resulting in an improvement of
the quantitative description when compared to experimental data, as will be
shown in the next section.

\section{Results}

In Figure 1 we show the relative viscosities predicted form the improved DEMT,
Eq.(\ref{viscosity1}), and Pal's model 2, Eq.(\ref{model2}), as a function of
the viscosity ratio $K$. Both models initially show a nearly constant value
followed by an increase in the range $10^{-1}<K<10^{3}$, approximately, and
finally becomes nearly constant again for large values of $K$. They also
predict an increase with increasing $\phi.$ In this figure, $\phi_{c}$ was
taken to be $0.7404$. A comparison between both models show that the improved
DEMT, Eq.(\ref{viscosity1}), predicts higher values of the relative viscosity
than Pal's model 2, Eq.(\ref{model2}), over the full range of $K.$ The
difference between both models is larger for larger $\phi$.

The various data sets considered for comparison with the models are the same
as the ones used in Ref. \cite{pal1}. They consist on emulsions covering a
viscosity ratio $3.87\times10^{-4}<K<3.25\times10^{5}$\ and with the
characteristic that the capillary number is small and that the emulsions are
stable (unfloculated).

Figure 2 shows comparisons between the experimental data and predictions of
improved DEMT, Eq.(\ref{viscosity1}), and Pal's model 2, Eq.(\ref{model2}),
for different values of the viscosity ratio $K$. As can be seen, improved DEMT
describes the experimental data slightly better than Pal's model 2. However,
the more significant difference between both models is the fitting parameter
$\phi_{c}$ which varies for different values of $K$. In all cases the improved
DEMT predicts larger values of $\phi_{c}$. Figure 3 shows the independence of
the relative viscosity on the mean droplet size. The comparison with the
theoretical models show again a better fit for the improved DEMT. In Fig. 4 we
compare experimental relative viscosity data at a fixed $\phi=0.5$ as a
function of the viscosity ratio $K$. Both models show a good agreement with
experimental data, nonetheless approximate fitting values of\ $\phi_{c}$ are
very different in both models, being larger for the improved DEMT.

According to Eq.(\ref{viscosity1}), the viscosity, plotted in the form
$\eta_{r}\left(  \phi\right)  ^{-2/5}\left[  (2\eta_{r}\left(  \phi\right)
+5K)/(2+5K)\right]  ^{-3/5}$ versus $\phi_{eff}$, should be system
independent. This is confirmed in Fig. 4 which shows all the viscosity data,
and compare them with the improved DEMT prediction. Pal's model 2 can not be
plotted in this sample independent form since it produces different curves for
different values of $\phi_{c}$. As can be seen, the agreement between data
points and improved DEMT is excellent.

\section{Conclusions}

An improved viscosity equation is developed for emulsions of spherical
droplets starting from the Taylor's equation for the viscosity of very dilute
emulsions \cite{taylor} and using a differential effective medium approach.
Geometrical information about the packing of the droplets is included through
an effective volume fraction, $\phi_{eff}$, which approaches $\phi$ at low
concentrations and becomes one at the critical concentration. The model
improves the results obtained with Pal's model 2, Ref. \cite{pal1} for the
relative viscosity, describes experimental data very well, and reduces to
Taylor's expression at low concentrations. In the limit $K\rightarrow\infty$
it reduces to the expression recently found for a suspension of hard spheres
\cite{mendoza}.

{\LARGE Acknowledgements}

This work was supported in part by Grant DGAPA IN-107607 (CIM) and DGAPA
IN-102609 (ISH).

{\LARGE Figure Captions}\bigskip

Fig. 1. Relative viscosity $\eta_{r}\left(  \phi\right)  \equiv\eta\left(
\phi\right)  /\eta_{0}$ predicted from the improved DEMT, Eq.(\ref{viscosity1}%
) (solid line) and Pal's model 2, Eq.(\ref{model2}) (dashed line). $\phi_{c}$
is taken to be $0.7404$.

Fig. 2. Comparisons between the experimental data [Pal (2001)] and best fits
of the theoretical predictions for the volume dependence of the relative
viscosity for different values of the viscosity ratio $K$. The solid line is
the result of the improved DEMT and the dashed line the prediction of Pal's
model 2.

Fig. 3. Comparisons between the experimental data and predictions of improved
DEMT (solid line), Eq.(\ref{viscosity1}) and Pal's model 2 (dashed line),
Eq.(\ref{model2}). The best fit for the improved DEMT was obtained with
$\phi_{c}=0.67$ and for Pal's model 2 $\phi_{c}=0.61$.

Fig. 4. Viscosity ratio ($K$) dependence of the relative viscosity. The solid
line is the result of the improved DEMT with $\phi_{c}=0.7404$ and the dashed
line is the prediction of Pal's model 2 with $\phi_{c}=0.63$.

Fig. 5. Master curve of $1-\phi_{eff}$ consisting of all the experimental
viscosity data. The solid line is the result of the improved DEMT.

\newpage%

\begin{figure}
[ptb]
\begin{center}
\includegraphics[
height=3.8553in,
width=5.028in
]%
{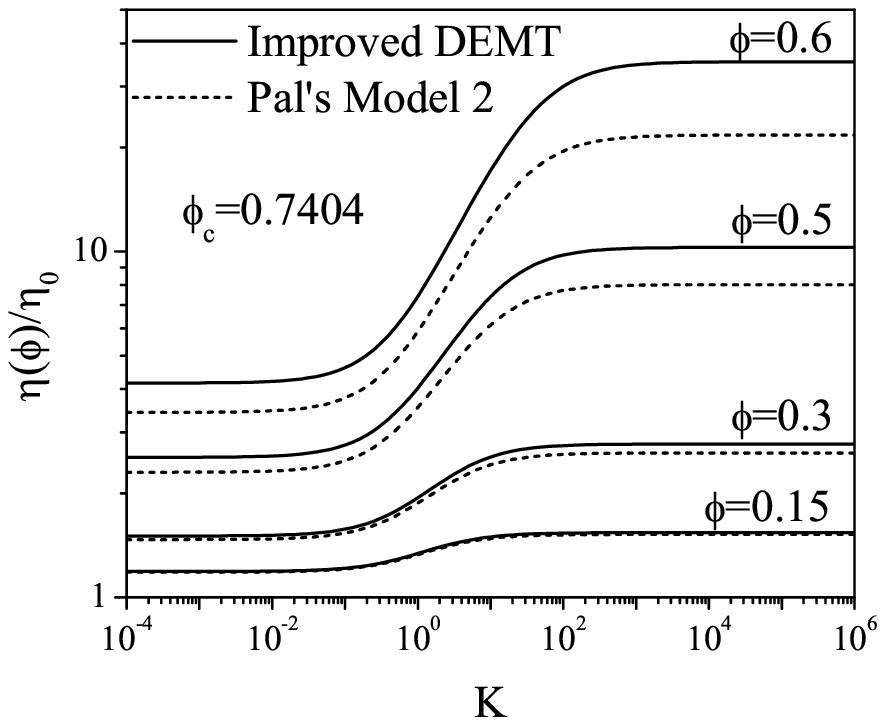}%
\end{center}
\caption{}
\end{figure}

\begin{figure}
[ptb]
\begin{center}
\includegraphics[
height=4.6345in,
width=6.0295in
]%
{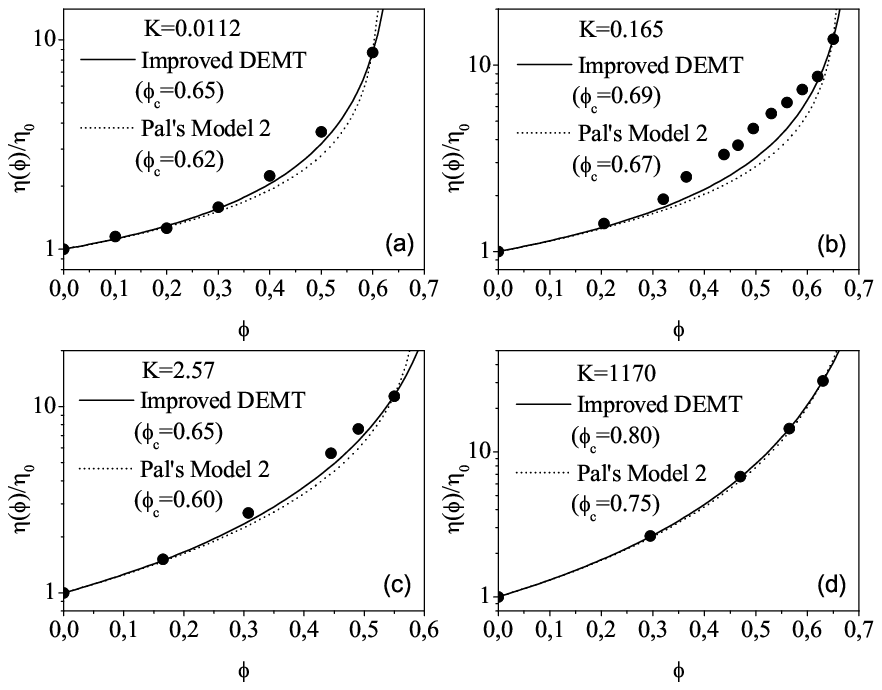}%
\end{center}
\caption{}
\end{figure}

\begin{figure}
[ptb]
\begin{center}
\includegraphics[
height=3.8553in,
width=5.028in
]%
{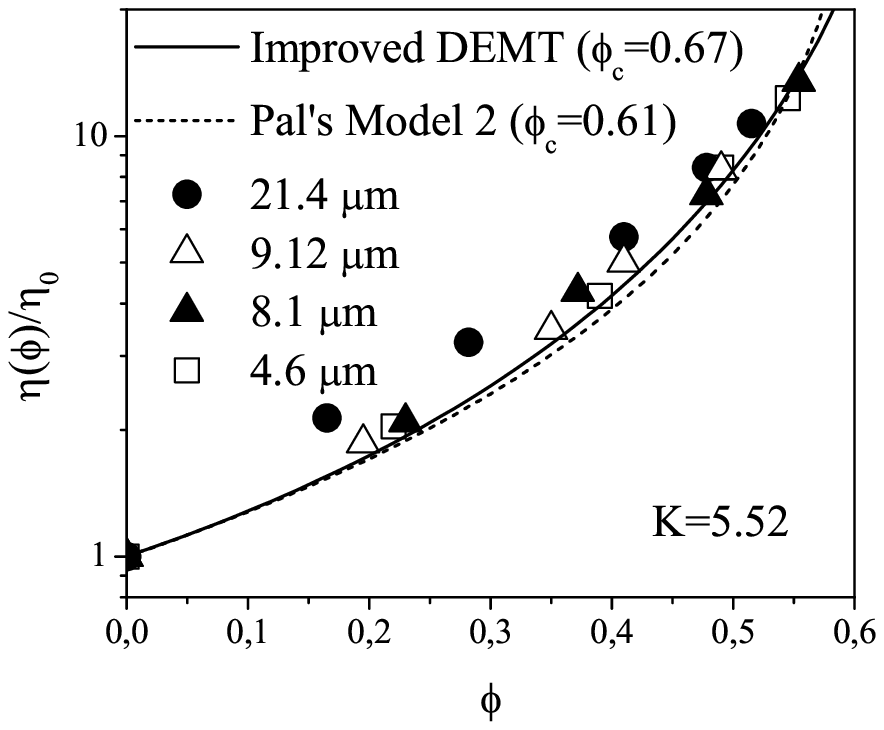}%
\end{center}
\caption{}
\end{figure}

\begin{figure}
[ptb]
\begin{center}
\includegraphics[
height=3.8553in,
width=5.028in
]%
{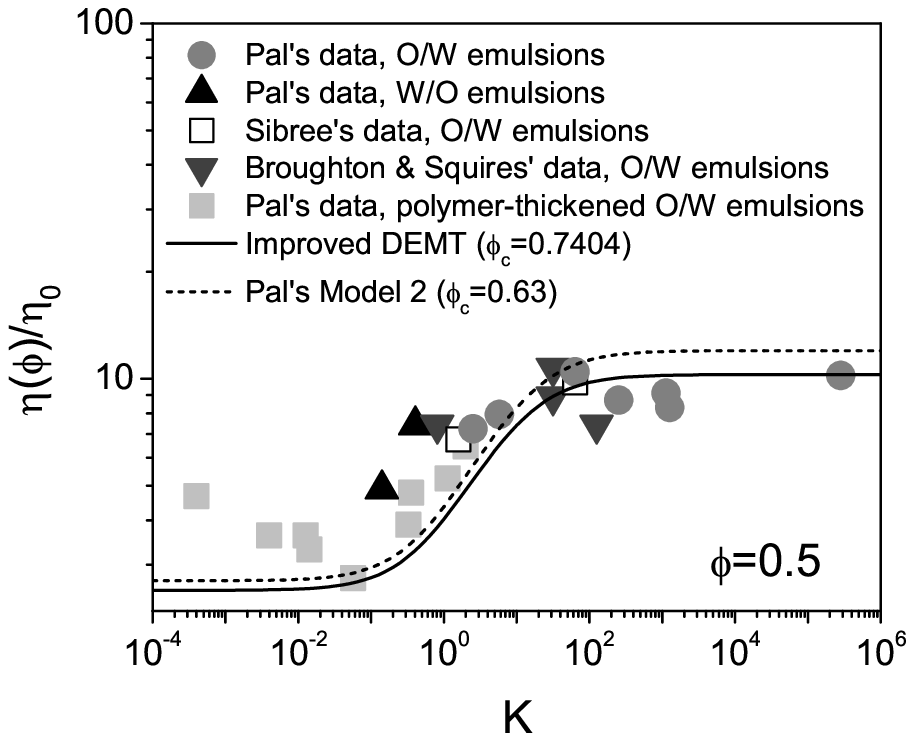}%
\end{center}
\caption{}
\end{figure}

\begin{figure}
[ptb]
\begin{center}
\includegraphics[
height=3.8553in,
width=5.028in
]%
{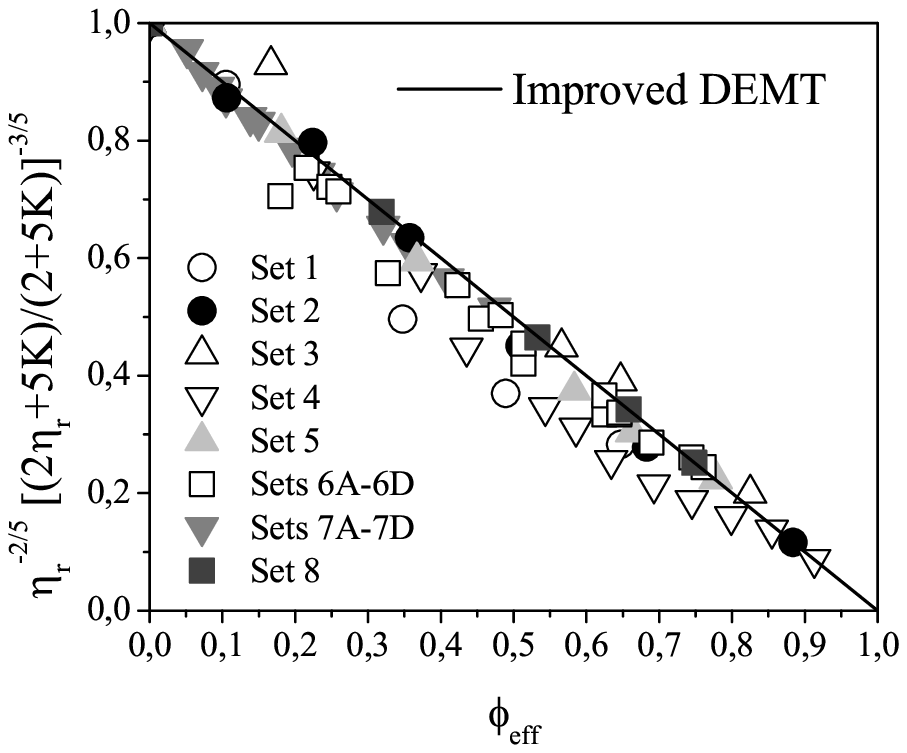}%
\end{center}
\caption{}
\end{figure}

\end{document}